\documentclass[onecolumn,preprintnumbers,amsmath,amssymb,superscriptaddress]{revtex4}
\usepackage{graphicx,subfigure,epsfig,epstopdf}
\usepackage{graphicx}
\usepackage{dcolumn}
\usepackage{bm}
\usepackage{amsmath,amssymb,upgreek}
\input{epsf}
\input{epsfx}

\makeatletter

\newcommand{\Rmnum}[1]{\expandafter\@slowromancap\romannumeral #1@}
\makeatother

\begin{document}
\title{Nanophotonic Filters and Integrated Networks in Flexible 2D Polymer Photonic Crystals}


\author{Xuetao Gan}
\affiliation{Department of Electrical Engineering, Columbia University, New York, NY 10027, USA}
\affiliation{School of Science, Northwestern Polytechnical University, Xi'an 710072, China}
\author{Hannah Clevenson}
\affiliation{Department of Electrical Engineering, Columbia University, New York, NY 10027, USA}
\affiliation{Department of Electrical Engineering and Computer Science, Massachusetts Institute of Technology, Cambridge, MA 02139, USA}
\author{Cheng-Chia Tsai}
\affiliation{Department of Electrical Engineering, Columbia University, New York, NY 10027, USA}
\author{Luozhou Li}
\affiliation{Department of Electrical Engineering, Columbia University, New York, NY 10027, USA}
\author{Dirk Englund}
\affiliation{Department of Electrical Engineering, Columbia University, New York, NY 10027, USA}
\affiliation{Department of Electrical Engineering and Computer Science, Massachusetts Institute of Technology, Cambridge, MA 02139, USA}
\affiliation{Dept. of Applied Physics, Columbia University, New York, NY 10027, USA}

\begin{abstract}
Polymers have appealing optical, biochemical, and mechanical qualities, including broadband transparency, ease of  functionalization,  and biocompatibility. However, their low refractive indices have precluded wavelength-scale optical confinement and nanophotonic applications in polymers. Here, we introduce a suspended polymer photonic crystal (SPPC) architecture that enables the implementation of nanophotonic structures typically limited to high-index materials. Using the SPPC platform, we demonstrate nanophotonic band-edge filters, waveguides, and nanocavities featuring quality ($Q$) factors exceeding $2,300$ and mode volumes ($V_{mode}$) below \textbf{$1.7(\lambda/n)^{3}$}. The unprecedentedly high $Q/V_{mode}$ ratio results in a 
 spectrally selective enhancement of radiative transitions of embedded emitters via the cavity Purcell effect with an enhancement factor exceeding 100. Moreover, the SPPC architecture allows straightforward integration of nanophotonic networks, shown here by  a waveguide-coupled cavity drop filter with sub-nanometer spectral resolution. The nanoscale optical confinement in polymer promises new applications ranging from optical communications to organic opto-electronics, and nanophotonic polymer sensors.
\end{abstract}
\maketitle


Photonic crystals (PCs) in semiconductors allow wavelength-scale control over the propagation of light and its interaction with matter, enabling precision sensors\cite{2007.OpticsExpress.Lee}, integrated on-chip networks\cite{McNab:03, Shinya:05}, classical\cite{2005.Science.Noda.SE_control,2008.LPR.Englund.Laser_review, Corcoran2009} and non-classical\cite{2005.PRL.Englund} light sources, and coupled quantum dot cavity systems for quantum information processing\cite{2004.Nature.Yoshie,2007.Nature.Hennessy-Imamoglu.Strong_coupling_quantum_nature,2007.Nature1}.
Because they leverage unique attributes of polymers, polymer-based PCs have inherent advantages for nanophotonics. For instance, in polymer PCs, the lattice structures as well as optical properties can be tuned easily through electrical, mechanical, and biochemical approaches\cite{Park2004,Zhao2012c}; various additives can be incorporated into polymer PCs to achieve multi-functional passive and active photonic devices\cite{Lodahl04,Zhao2012c}. While polymer PCs in different dimensions have been fabricated using various methods\cite{Lodahl04,Choi2012a,Wulbern2007,Wu2007,Shen2012,Zhang2012g}, the designs reported to date have not been suitable for integration of multiple devices into PC networks. Although polymer PCs could in principle match the optical confinement of their semiconductor membrane counterparts, a typical transparent polymer has a low refractive index 
($n\sim1.5$). This would require designs that simultaneously achieve the required in-plane photonic bandgap (PBG) and total internal reflection (TIR) for all relevant $k-$states constituting the resonant and traveling modes employed in the planar photonic crystals (PPCs). To obtain a PBG in polymer PPCs, several approaches that employ low refractive index ($n\sim1.15$) substrate or suspend the polymer film in air have been proposed\cite{Choi2012a,Wulbern2007,Wu2007}.  However, experimental realisations of polymer PPC defect modes due to the confinement of the PBG are still inconclusive. In this paper, we describe a simple one-step lithography technique for the fabrication of a new generation of suspended polymer photonic crystals (SPPCs) with optimised vertical TIR and with transverse field confinement achieved by a complete PBG for all in-plane transverse electric (TE)-like wave components. Notably, this SPPC architecture allows for wavelength-scale optical confinement in a polymer PC nanocavity with nearly an order of magnitude smaller mode volume than previous polymer cavities\cite{Quan:11, Ling:11, Kim:09}, as well as for nanophotonic waveguides and straightforward integration of multiple devices into planar networks.

It has been shown before that the high Purcell effect in semiconductor nanocavities can greatly increase coupling between emitters and the cavity mode, which has been used for numerous advances including controlled spontaneous emission (SE)~\cite{2005.PRL.Englund,2005.Science.Noda.SE_control}, strong emitter-photon coupling,~\cite{2004.Nature.Yoshie,2007.Nature.Hennessy-Imamoglu.Strong_coupling_quantum_nature,2007.Nature1} and laser oscillation at ultra-low threshold~\cite{2006.PRL.Strauf}. The exceptionally high $Q/V_{mode}$ ratio of polymer nanocavities demonstrated here offers the prospect of strong SE rate modification in polymer devices. In the experiments described below, we use organic dye (Coumarin 6) dispersed throughout the polymethyl-methacrylate (PMMA) film as the quantum emitters. With a largely homogeneous emission spectrum~\cite{Prahl} much broader than that of the cavity resonant peak, we expect that for a dye molecule in the cavity at position $\mathbf{r}$, the decay rate of the excited state $1/\tau$ is related to the sum of decay rates into nonradiative and radiative transitions
\begin{equation}\label{eq:purcell}
\frac{1}{\tau}=\int\mathrm{d}\lambda \gamma_0(\lambda) [\eta_{PC}F_{PC}(\mathbf{r},\lambda)+\eta_{cav} F_c(\lambda,\mathbf{r})]+\Gamma_{NR},
\end{equation}
where $\gamma_0(\lambda)d\lambda$ refers to the SE rate of the dye molecule in unpatterned PMMA into the wavelength range ($\lambda,\lambda+d\lambda$), $\Gamma_{NR}$ is the nonradiative recombination rate, and the factors $F_{PC}(\lambda,\mathbf{r})$ and $F_c(\lambda,\mathbf{r})$ correspond to the modifications of the SE rate when the emitter excited state is coupled to the  PC leaky mode and the cavity mode, respectively\cite{2005.PRL.Englund}. Because of the small bandgap in the SPPC, the suppression of the cavity-enhanced emission rate from $F_{PC}$ is weak and we estimate $F_{PC}\sim 1$. The term $F_c(\lambda,\mathbf{r})=F_{c,0}L(\lambda)\psi(\mathbf{r})^{2}$ describes the modification of the SE by the cavity mode, where  $F_{c,0}=\frac{3}{4\pi^{2}}\frac{Q}{V_{mode}}(\frac{\lambda_{cav}}{n})^3$ is the maximum SE rate enhancement (Purcell) factor, $L(\lambda)=1/[1+4Q^2(\frac{\lambda}{\lambda_{cav}}-1)^2]$ is the cavity lineshape function, and $\psi(\mathbf{r})=\frac{\mathbf{E(\mathbf{r})}\cdot\mathbf{\upmu}(\mathbf{r})}{|\mathbf{E(\mathbf{r})}_{max}||\mathbf{\upmu}(\mathbf{r})|}$ denotes the spatial alignment of the dye molecule's electric dipole moment $\mathbf{\upmu}(\mathbf{r})$  to the cavity field $\mathbf{E}(\mathbf{r})$. The factors $\eta_{PC}$ and $\eta_{cav}$ correspond to the collection efficiencies into the microscope mode from the averaged PC leaky modes  and the cavity mode, respectively. When the emitter linewidth is larger than the cavity linewidth (the `bad-emitter' regime), as is the case for the broad homogenous linewidth of Coumarin 6 compared to the narrow SPPC cavity linewidth, the enhancement of the SE into a small subset of optical transitions can cause a strong modification of the emission spectrum. However, because of the narrow cavity linewidth, the contribution of the cavity term in the integral in Eq. \ref{eq:purcell} is small; thus, the excited state lifetime may be reduced only slightly.

\noindent\textbf{Result}

While a wide range of polymers may be used, we employ here PMMA, which is an electron beam lithography (EBL) resist with a wide optical transparency spanning the visible to infrared spectrum. We fabricate SPPC devices using EBL on a $\sim$ 400~nm thick PMMA and polyvinyl alcohol (PVA) bilayer (see Methods). The water-soluble PVA spacer layer allows for lift-off and transfer of the developed PMMA film onto various carrier substrates, such as a $\sim$cm$^{2}$-gap in a flexible polyester carrier (Fig. 1a), an organic spun fiber tissue (Fig. 1b), or a flexible polymer substrate (Fig. 1c). The adhesion to the carriers is strong enough to hold the whole film robustly. As shown in Fig. 1d, the finished SPPC devices are free of wrinkles and distortions. Since this fabrication process eliminates the typical mask-transfer steps used for semiconductor PCs, the  SPPC devices show excellent spatial resolution free of defects, as seen in scanning electron microscope (SEM) images of a photonic drop filter network (Fig. 1e) and a  band-edge filter (Fig. 1f).

 \begin{figure}[th!]\centering
\includegraphics[width=6in]{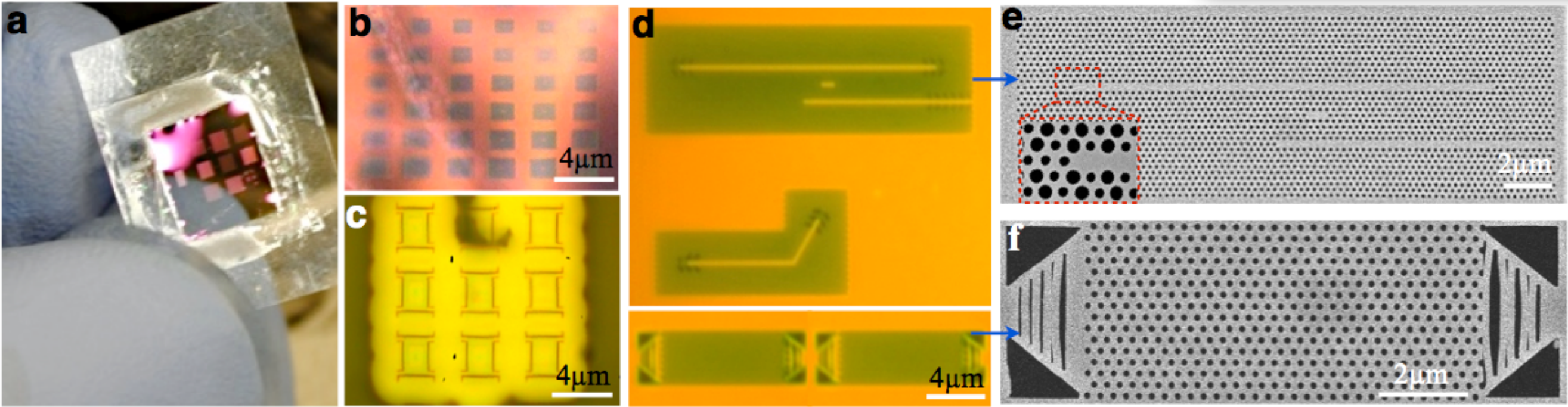}
\caption{{\small {\bf{Images of the SPPC devices.}} {\bf{a}}, Photograph of the cm$^2$-sized PMMA film with SPPC devices mounted on a flexible polyester carrier. {\bf{b,c}}, Optical microscope images of the SPPC devices transferred onto the fiber tissue and photoresist polymer (MICROPOSIT S1811) substrate after wet undercutting. {\bf{d}}, Optical microscope image of the tested SPPC devices, where from top to bottom show SPPC drop-filter, bent waveguide, and band-edge filter along $\Gamma$X and $\Gamma$J direction. {\bf{e,f}}, Scanning electron microscope images of the SPPC drop-filter and band-edge filter.  } }
\label{fig:equipment_layout} 
\end{figure}
\vspace{12pt}

We design the SPPC devices with three-dimensional (3D) finite-difference time-domain (FDTD) simulations (see Method).  Fig. 2a shows the simulated band diagram for a suspended 2D triangular hole-array. From a PC design with lattice spacing $a$, hole radius $r=0.3a$, slab thickness $d=1.3a$, and index of refraction $n=1.52$ for our PMMA film (measured by ellipsometry), we simulate a PBG with frequencies between $0.487(2{\pi}c/a)$ to $0.503(2{\pi}c/a)$, where $c$ is the speed of light in vacuum. The membrane thickness is optimised for single mode behavior and a high in-plane bandgap, allowing for 3D light confinement by both in-plane distributed Bragg reflection and vertical TIR. 

 \begin{figure}[th!]\centering
\includegraphics[width=6in]{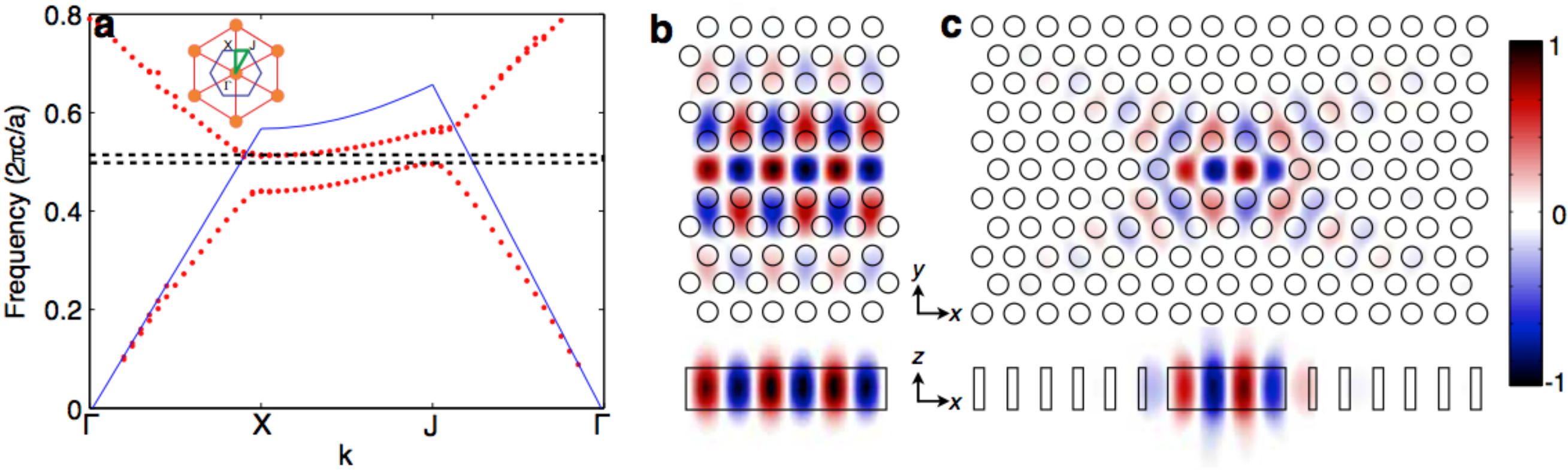}
\caption{{\small {\bf{Simulations of the triangular lattice SPPC devices with parameters $r/a=0.3$, $d/a=1.3$, where $a$ is the lattice spacing.}} {\bf{a}}, TE-like mode band diagram for the SPPC with the bandgap between  $0.487(2{\pi}c/a)$ and $0.503(2{\pi}c/a)$. The blue solid line displays the light line for air. The inset shows the reciprocal lattice (orange circles), the first Brillouin zone (blue) and the irreducible Brillouin zone (green) with the high symmetry points. {\bf{b}}, Guided mode ($H_z$-component) of the SPPC waveguide at $k=\pi/a$ in plane (top) and in cross-section (bottom), which illustrates the well confinement of light in the sub-wavelength scale. {\bf{c}}, Confined fundamental modes  ($H_z$-component) of the L3 nanocavity in plane (top) and in cross-section (bottom). The $Q$ factor and the mode volume of the resonant mode are ~3,000 and $1.68(\lambda/n)^3$, respectively. } }
\end{figure}
\vspace{12pt}

By introducing a single-line defect along the $\Gamma$J direction, light is confined in a manner similar to guided modes in high-index PCs, as shown in Fig. 2b. Terminating the waveguide around three missing holes and optimising the local geometry\cite{Noda2003Nature} yields a linear three-hole (L3) cavity representing a fundamental mode with $Q\sim 3,000$ at a frequency of $0.491(2\pi{c}/a)$. Due to the low refractive index of the polymer, the cavity mode penetrates deep into the lattice and has a strong evanescent field outside of the slab (Fig. 2c). The normalised mode volume of 1.68 $(\lambda/n)^3$ is only twice as large as that of a silicon L3 cavity\cite{2005.OpEx.Noda.highQ} and, to our knowledge, represents an order of magnitude reduction in the mode volume of previously demonstrated polymer cavities\cite{Quan:11, Ling:11, Kim:09}. We find in a hetero-waveguide cavity that the $Q$ factor can be even greater, reaching a simulated $Q$ factor as high as 19,000 due to the gradual mode-matching in the cavity\cite{Gai2012,2005.NMat.Akahane} (see Supplementary Information).

We characterize the fabricated SPPC devices with a confocal microscope (see Methods). An organic dye (Coumarin 6, $5\%$ by weight) dispersed in PMMA acts as the coupled emitter system and also provides a convenient internal light source for characterisation of the optical properties of the devices. In our experiments, the dye is excited by a 405~nm continuous-wave  laser and emits from 480~nm to 650~nm. To characterise the PBG of SPPC,  we perform transmission measurements through band-edge filters such as the one shown in Fig. 1f, which  have lattice parameters $a= 300$~nm, $r=90$~nm, and linear grating couplers at both ends for vertical coupling. In these measurements, we excite broadband photoluminescence (PL) in one of the grating couplers and collect the transmission in the other (right panel, Fig. 3a). The left panel of Fig. 3a displays the transmission spectra along the $\Gamma$X and $\Gamma$J directions. The transmitted intensities along the two directions show a clear common stop band between $\lambda= 597-615$~nm, corresponding to  normalised frequencies of $a/\lambda=0.487-0.503$. These results precisely match the above-mentioned simulations. 

With this unambiguous demonstration of the  PBG, we characterise the L3 SPPC nanocavities and the dye's coupling by the emitter-cavity emission spectra shown in Fig. 3b.  The plots show several linearly polarised cavity peaks with orthogonal polarisations. Fitting the fundamental mode at $611.3$~nm to a Lorentzian yields a high $Q$ of $2,300$. The cavity PL image for the fundamental mode, shown in the inset of Fig. 3b, displays two bright lobes at the ends of the L3 cavity (top), matching the simulation of the far-field radiation pattern of the mode (bottom).
By comparing the PL spectra at different polarisations, we observe remarkably strong modifications of the dye's emission spectrum, which indicates that the polymer cavity-dye system effectively represents a hybrid emitter whose emission spectrum is much narrower than that of dye molecules alone. By controlling the dye's SE properties in the high $Q/V_{mode}$ polymer cavity, it is possible to engineer extremely narrow emission properties of dyes or other emitters with high homogenous broadening.
 
 \begin{figure}[th!]\centering
\includegraphics[width=6in]{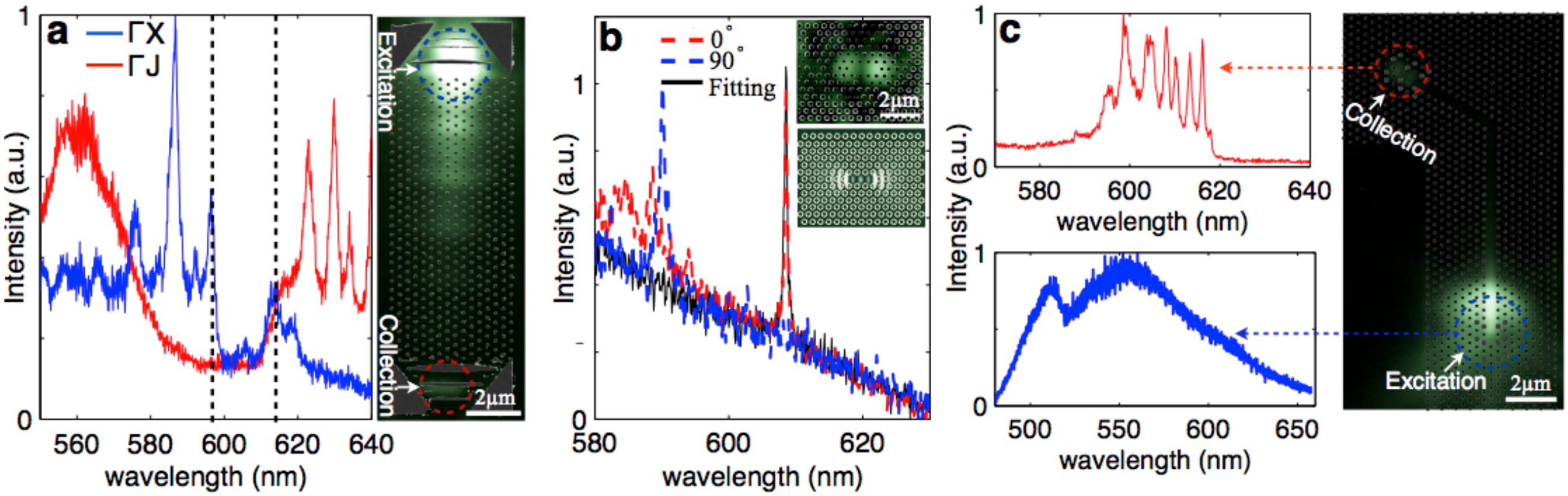}
\caption{{\small {\bf{Experimental characterisation results of the SPPC bandgap, L3 cavity resonance, and the transmission of the bent waveguide.}} {\bf{a}}, left: Transmission spectra of the SPPC band-edge filters along $\Gamma$X and $\Gamma$J directions. The region between two dashed lines represents the in-plane complete bandgap in the spectral range of 597-615~nm. right: PL image of the band-edge filter superimposed with the SEM image. Excited on one side, the PL due to the dye transmits through the SPPC filter and is collected on the other grating coupler. {\bf{b}}, PL spectra collected from the L3 cavity, which indicate sharp cavity resonant modes with orthogonal polarisations (dashed lines).  The fundamental mode at 611.3~nm has a $Q$ factor of 2,300 fitting to the Lorentzian. The black line shows the fitting result from the spectrally selective SE rate enhancement of optical transitions coupled to the fundamental mode. The inset shows the PL image of the fundamental cavity mode superimposed with the cavity SEM (top), which has two bright lobes at the terminals of the defect; the bottom one is the simulation of the far field of $|E_y|^2$.  {\bf{c}}, left: Spectra collected from the excitation (bottom) and collection (top) ports of the $60^\circ$ bent waveguide, where the narrow transmission indicates the waveguide pass-band. right: PL image of the bent waveguide superimposed with the SEM image, presenting a bright spot at the output grating coupler.  } }
\end{figure}
\vspace{12pt}

Next, we describe how to integrate isolated nanophotonic components into networks on the SPPC architecture. Shown in the center panel of Fig. 1d is an SPPC waveguide with a 60$^\circ$ bend --- important network components that have not previously been demonstrated in polymer. As in a high-index membrane, the waveguide consists of a line defect in the SPPC. Enlarged air-holes at the ends of the waveguide function as vertical couplers\cite{2011.OpEx.WG_coupler} and therefore appear dark in the optical micrograph. The enlarged air-holes of the coupler, as shown in the inset of Fig. 1e,  have a period of 2$a$, which couples the confined modes near the $k_x=\pi/a$ point to the radiative light cone to improve the vertical coupling of the waveguide mode\cite{Encoupler}. The radius of the enlarged air-holes is set at 0.4$a$.  Fig. 3c displays the transmission of a broad-band excitation field, formed again by dye fluorescence at the waveguide input. The PL image (Fig. 3c, right) shows the transmitted light scattering out at the waveguide output coupler. The spectra collected from the excitation (bottom left) and collection (top left) couplers show a pass-band from 597~nm to 617~nm. The Fabry-Perot (FP) oscillations in the spectra are due to etalon effects from imperfectly matched waveguide terminations, and can be eliminated by replacing the terminated air-holes with an extended waveguide.  Note that the FP-type resonant peaks only enhance the light reflection at certain wavelengths, and have no influence on the characterisation of the waveguide transmission band. By designing SPPC waveguides with different lengths and measuring the extinct ratio of the oscillation peaks\cite{WALKKER}, we obtain the transmission loss of the waveguide  close to 80~dB/cm over the transmission band. The large transmission loss is attributed to a lack of uniformity caused by the aberration of the EBL writing over large length-scales using a converted SEM-based EBL system without interferometric stages.
From the simulations on the guiding mode shown in Fig. 2b, we estimate the transmission loss can be lower than 1 dB/cm. We anticipate that devices with much lower losses are possible using high-voltage EBL systems with stitching capability.

The SPPC waveguides enable complex networks.  Fig. 4a shows a simulation of a channel drop filter consisting of two waveguides coupled via an L3 nanocavity, which is separated from the waveguides by four rows of holes to achieve nearly critical coupling. To simulate the network's operation, we excite the input waveguide with a spectrally broad pulse. The output port  shows narrow spectral peaks that correspond to the cavity resonances.

The fabricated SPPC channel drop filter includes vertical grating couplers at the ends of the waveguides\cite{2011.OpEx.WG_coupler,Encoupler}, as shown in Fig. 1e. A broadband input field is generated by pumping the input waveguide and transmits through the device, as shown in Fig. 4b. Fig. 4c plots spectra collected from (i) the top grating of the input waveguide, (ii) the cavity, and (iii) the bottom grating of the drop waveguide. FP-like oscillations in (i) result from the termination of the input waveguide, effectively forming a large, low-finesse cavity. The spectrum from the cavity mode at position (ii) shows a Lorentzian peak with a 0.52~nm linewidth. The spectrum from position (iii) plots the filtered intensity, showing a strong peak at the cavity resonance and sharply suppressed intensity at other spectral components, corresponding to an extinction ratio of 8~dB. The isolation of this drop filter is only slightly lower than the 12~dB extinction for silicon-based devices\cite{Takano:06}. The $Q$ factor of the cavity matches our simulations and may be increased significantly by extending the cavity length or by changing the coupling rate to the waveguides.

 \begin{figure}[th!]\centering
\includegraphics[width=6in]{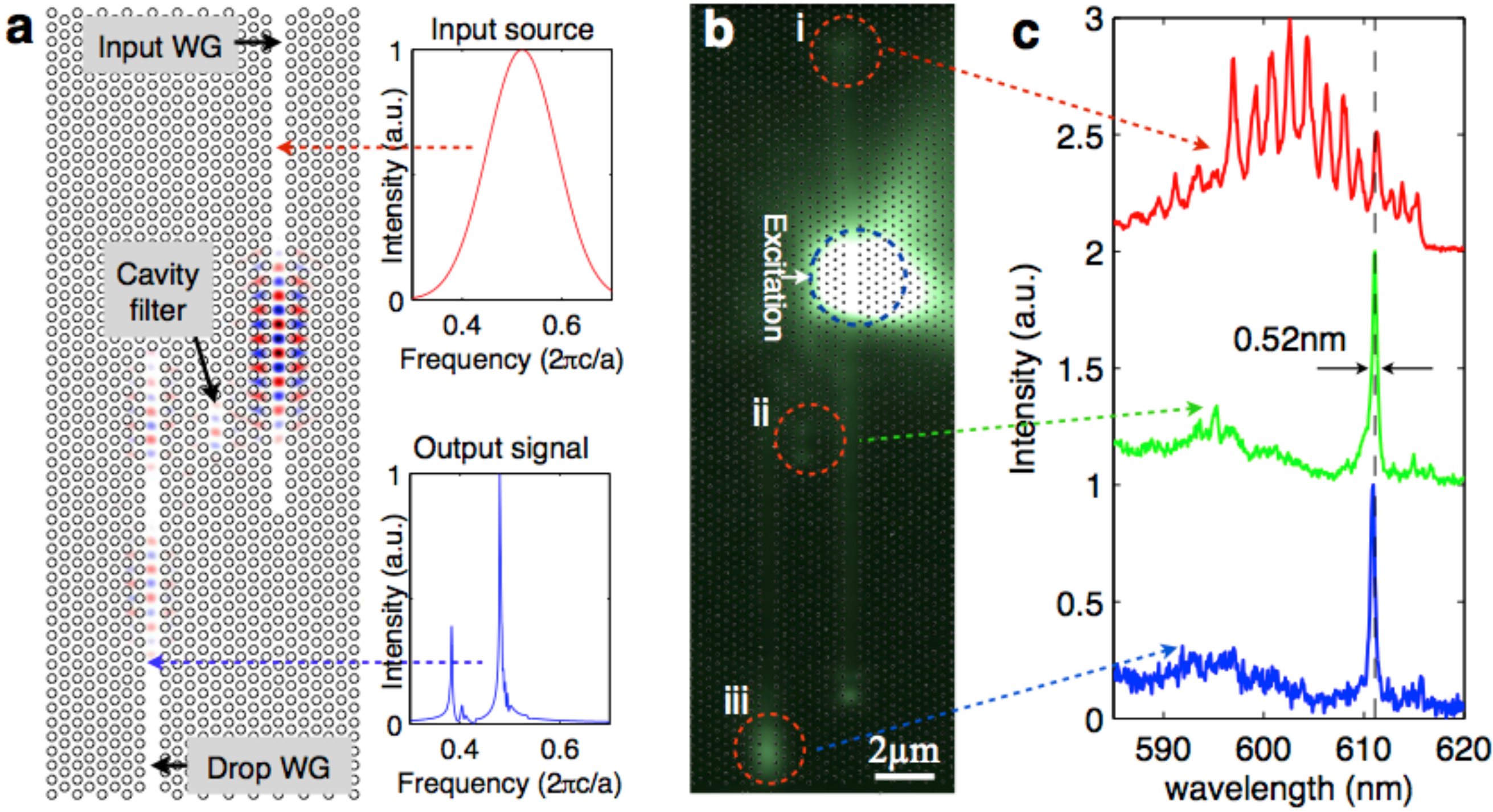}
\caption{{\small {\bf{Demonstration of the SPPC waveguide-coupled cavity drop filter consists of two waveguides coupled via a L3 nanocavity.}} {\bf{a}}, Simulation of the drop filter. A broadband source excites on the input waveguide, and is filtered by the cavity. We obtain a well-confined waveguide mode in the drop waveguide, which has a narrow spectral linewidth (right panel). {\bf{b}}, PL image of the drop filter device superimposed with its SEM image. The PL generated on the input waveguide transmits through the cavity filter and scatters out from the grating coupler on the waveguides. {\bf{c}}, Spectra of the light collected from (i) the top grating, (ii) the cavity, and (iii) the bottom grating of the drop waveguide. The peak on the spectra from (iii) indicates the sub-nanometer filtering of the cavity.   } }
\end{figure}
\vspace{12pt}

\noindent\textbf{Discussion} 

We develop an SPPC platform for integrated nanophotonic devices in polymer films, including band-edge filters, waveguides, and high-$Q$ nanocavities. The appealing properties of polymers enable new possibilities for planar photonic crystals. The implementation of the drop filter network indicates the SPPC architecture could serve as a passive architecture for on-chip routing. While the fabricated SPPC devices are characterized via the internal dye PL, optimized grating couplers can be designed to ensure efficient coupling of the device via the fiber-based vertical or edge butt-coupling. Based on our technique, which allows the fabrication of large-sized suspended polymer films (shown in Fig. 1a), it is possible to achieve the photonic circuits in large scale and build chip-to-chip and board-to-board optical communication in polymers, in line with next-generation optical interconnect technologies\cite{1707717}. Actively, the SPPC networks can be controlled by mechanical or electro-optic modulation, which may be as fast as 40 GHz in electro-optical polymers\cite{Wulbern:10}. 

The SPPC L3 cavity presenting a resonant mode with  $Q$ factor high as 2,300 and $V_{mode}$ low as \textbf{$1.68(\lambda/n)^{3}$} enables an unprecedentedly high Purcell factor $F_{c,0}=\frac{3}{4\pi^{2}}\frac{Q}{V_{mode}}(\frac{\lambda_{c,0}}{n})^3\approx 104$, which can dramatically modify the emission spectral properties of an organic emitter coupled to the high $Q/V_{mode}$ cavity mode. Applying Eq.~\ref{eq:purcell} across dye molecules dispersed uniformly throughout the polymer membrane, a spatial integral over emitter positions and random dipole orientations (in the solid angle) enables us to estimate the modification of the dye molecules' photoluminescence spectrum when exciting and collecting from a Gaussian spot focused on the cavity with a diameter of $300$~nm  (see Supplementary Information)\cite{2005.APL.Fushman.PbS,VanderSar2011}. This calculation predicts the emission spectrum shown in the black line of Fig. 3(b), assuming the maximum emission rate modification of $F_{c,0}=104$. This prediction agrees closely with the experimentally observed dye emission spectrum and indicates that emitters coupled to the cavity indeed experience a dramatic change in the emission spectrum, with the rate of optical transitions coupled to the cavity mode enhanced by as much as two orders of magnitude. Thus, through the spectrally selective enhancement of the SE rate into cavity-coupled transitions, a strong redistribution of the dye emission spectrum occurs. The ability to amplify the SE rate of spectrally selectable transitions by 100 or more opens up new possibilities in  solid state cavity quantum electrodynamics  applications with numerous other emitters, such as quantum dots, nitrogen vacancy center in nanocystals, and molecules\cite{Tischler:2005ys}. Here, the observed spectrally selective SE rate modification is lowered due to the spatial averaging over emitters across the collection region from the cavity. We anticipate that other devices, such as photonic crystal coupled cavity arrays\cite{2005.APL.Altug}, or spatially selective dye placement in the high-cavity field region only\cite{Rivoire2009}, could be used to realize a strong spontaneous emission rate modification across all emitters. 

Based on our proposed transfer technique of thin polymer films, multifunctional devices can be implemented on a variety of carriers, including roughened or patterned surfaces that provide an air-gap separation. The all-polymer spacer structure, shown in Fig. 1c, can be conformally transferred with high precision onto a wide range of surfaces to integrate nanophotonic structures and complex optical networks on a variety of materials and hybrid devices. The success of transferring and suspending sub-micron thin polymer films has great potential applications for other nanofabrication techniques.

Fabrication imperfections can shift the resonant wavelength of PC nanocavities strongly\cite{Nodahetero}, making it difficult to control the cavity resonance. However, for the SPPC nanocavities, 
 the characterisation of hundreds of devices with the same design shows that $98\%$ of cavity resonances fall within 1~nm of the center of the distribution. This implies low fabrication deviations in the single lithography-step fabrication technique. We also have tested the lifetime and stability of our fabricated SPPC devices, which were stored in the lab environment for one month. We have found no observable change in the resonance of the L3 cavity, as confirmed by repeated characterisation of more than one hundred devices. Fabrication technologies such as nanoimprint\cite{2007.Nanotech.Kim-Park} and direct laser writing\cite{Deubel:2004uq} can be used to make SPPC devices in large quantities. We expect that the SPPC platform will have applications in organic light emitters, organic photovoltaics\cite{Li:2012fk}, and other areas of organic opto-electronics\cite{Peyghambarian:05}.

 \vspace{1cm}
 \noindent{\bf{Supplementary Information:}}
 
\noindent{\bf{Device fabrication.}} To construct SPPC devices, we created a bilayer structure by spin coating 10~nm thick PVA and 390~nm thick PMMA layers on a bare silicon wafer. The PC patterns were defined through EBL and developed in a MIBK:IPA 1:3 solution. The patterned PMMA layer was dislodged from the wafer and floated to the  water surface where it could be transferred onto a carrier.

\noindent{\bf{Simulations of SPPC devices.}} The simulations of the SPPC devices were performed using a 3D finite-difference time-domain technique (MEEP). The air-hole radii and the slab thickness were normalised by the lattice constant $a$, and the spatial resolution was set at 1/30 of the lattice constant. The photonic band diagram was obtained by calculating the transmission bands of a supercell of the PC lattice along different $k$-vectors in the momentum space. Guided modes of the SSPC waveguide were simulated at $k=\pi/a$. The L3 cavity was formed by three linearly  missing holes and the two end holes were displaced along the cavity axis by 0.15$a$ in order to increase the $Q$-factor. 

\noindent {\bf{Sample characterization.}} The dye-doped PMMA film's thickness and refractive index were measured by an ellipsometer. The Coumarin 6 dye acted as an internal light source and generated PL spectra between $480-650$~nm when pumped with a $405$~nm continuous wave laser. SPPC devices were characterised using a confocal microscope with independent pump and collection paths, which could be manipulated using steering mirrors. A polariser placed in the collection path was used to analyse the polarisation dependence of the SPPC device signals, which are further analysed with a $0.5$~m spectrometer and a cooled silicon CCD. Transmission measurements of a SPPC device (such as the band-edge filters and bent waveguides) were performed using the steering mirrors to focus the pump and collection paths on different ends of the device. The L3 cavity was characterised with both the pump laser and collection spot on the defect region. The integrated network was tested by fixing the pump on the input waveguide while moving the collection spot onto (i), (ii), or (iii).

 \vspace{0.5cm}
 \noindent{\bf{ Acknowledgment:}}
 
Financial support was provided by the Air Force Office of Scientific Research PECASE, supervised by Dr. Gernot Pomrenke. FDTD simulations were carried out in part at the Center for Functional Nanomaterials, Brookhaven National Laboratory, which is supported by the U.S. Department of Energy, Office of Basic Energy Sciences, under Contract No. DE-AC02-98CH10886. X. G. was supported in part by the 973 program (2012CB921900). H.C. was supported in part by the NASA Space Technology Research Fellowship.

 \vspace{0.5cm}
 \noindent{\bf{ Author contributions: }}

D.E. and X.G. conceived the project.  X.G. designed, fabricated, and experimentally charactarised the devices. X.G., H.C., and D.E. prepared the manuscript. C.T. and L.L. did the simulations. All authors reviewed the manuscript.

 \vspace{0.5cm}
 \noindent{\bf{ Correspondence:  }}

Correspondence and requests for materials
should be addressed to D. E. (email: englund@mit.edu).



\end{document}